\documentclass[%
reprint,
 amsmath,amssymb,
 aps,showkeys,
prl
]{revtex4-1}

\usepackage[english,russian]{babel}
\usepackage{graphics}
\usepackage{graphicx}% Include figure files
\usepackage{dcolumn}% Align table columns on decimal point
\usepackage{bm}% bold math
\usepackage{hyperref}% add hypertext capabilities
\usepackage{placeins}
\usepackage{xcolor}
\usepackage[english,russian]{babel}

\begin{document}

%\thispagestyle{empty}

%\fontsize{14}{16pt}\selectfont

\title{
On the universal properties of stochastic processes under optimally tuned Poisson restart}
%\date\today

\author{Sergey Belan}
\email{sergb27@yandex.ru}

\affiliation{$^{1}$Landau Institute for Theoretical Physics, Russian Academy of Sciences, 1-A Akademika Semenova av., 142432 Chernogolovka, Russia}
\affiliation{$^{2}$National Research University Higher School of Economics, Faculty of Mathematics, Usacheva 6, 119048 Moscow, Russia}
\affiliation{$^{3}$National Research University Higher School of Economics, Faculty of Physics, Myasnitskaya 20, 101000 Moscow, Russia}

\begin{abstract} 
Poisson restart assumes that a stochastic process is interrupted and starts again at random time moments.
A number of studies have demonstrated that this strategy may minimize the expected completion time in some classes of random search tasks.
What is more, it turned out that under optimally tuned restart rate, any stochastic process, regardless of its nature and statistical details, satisfies a number of universal relations for the statistical moments of completion time.
In this paper, we describe several new universal properties of optimally restarted processes.
Also we obtain a universal inequality for the quadratic statistical moments of completion time in the optimization problem where 
stochastic process has several possible completion scenarios. 
\end{abstract}

\maketitle

\vskip \baselineskip

\textit{Introduction.}
Restarting - the abrupt interruption of a stochastic process followed by the initiation of a new statistically independent attempt - was proposed several decades ago in computer science as a method for increasing the performance of probabilistic algorithms  \cite{Alt_1991, Luby_1993}.
It turned out that if the fluctuations in the completion time of the randomized algorithm are large enough, then restarts at a properly selected frequency improves (on average) performance. This effect has a fairly simple qualitative explanation: if the current implementation of a randomized algorithm takes too long, then it may be  more profitable to abort it and try again than to wait for the program to terminate. 

%On universal properties of random processes under optimal Poisson restart

Later, the potential for increasing the speed of a random process by restarting  was discovered in the enzymatic kinetics.
An elementary act of an enzymatic reaction begins with the reversible binding of a substrate to an enzyme.
If the intermediate enzyme-substrate is formed, some time is required for the catalytic step, during which the initial substrate is converted into products.
However, the reversibility of enzyme-substrate binding means that the intermediate complex can dissociate before the products are formed.
In this case, the next attempt of catalysis can only take place after the enzyme will find another substrate.
In other words, the dissociation of the intermediate enzyme-substrate corresponds to restart of the enzymatic reaction.
In theoretical works \cite{Reuveni_2014, Rotbart_2015, Reuveni_PRL_2016,berezhkovskii2017bulk} it was shown that if the intermediate complex has several fairly long-lived conformations characterized by significantly different rates of formation of reaction products, then an increase in the rate of substrate unbinding may lead to growth of  the reaction rate. 

Depending on the context, it may be interesting to explore different restart protocols.
Say, in computer science, it is often assumed that a randomized algorithm restarts according to some predetermined schedule \cite{Wu_2007,Lorenz_2018,Lorenz_2021,schulte2010modeling, cire2014parallel, amadini2018sunny,van2006analysis,wallace2020search,schroeder2001does}.
On the contary, in the case of chemical transformations, restart events occur at random time moments since decay of the intermediate complex enzyme-substrate happens due to thermal fluctuations, see e.g. \cite{Reuveni_2014, Rotbart_2015, Reuveni_PRL_2016,berezhkovskii2017bulk}.

Among many posible restart strategies, stochastic restart at Poisson times is most extensively studied in the existing literature, see, e.g.,  \cite{EM_2011,evans2020stochastic,evans2011diffusion,evans2014diffusion,evans2018run,blumer2024combining,julian2024diffusion,Pal_JPA_2022,Pal_2019a,bonomo2021first,ray2020diffusion,ray2019peclet}.
The average value of the random completion time $T_r$ of a process, restarted with a rate $r$, is given by a simple formula \cite{Reuveni_PRL_2016}
\begin{equation}\label{eq:T_r_mean_main}
\langle T_r\rangle=\frac{1-\tilde{P}(r)}{r\tilde{P}(r)},
\end{equation}
where $\tilde{P}(r)=\int_0^\infty dT P(T)e^{-rT}$ is the Laplace transform of the completion time probability density $P(T)$ in the absence of restart.

The work of Evans and Majumdar \cite{EM_2011} showed that there is an optimal restart rate $r_\ast>0$ that minimizes the average completion time of one-dimensional diffusion-mediated search.
Following this, the possibility of similar optimization was demonstrated for many other random search problems \cite{evans2011diffusion,whitehouse2013effect,evans2014diffusion,kusmierz2014first,kusmierz2015optimal,evans2018run,ray2019peclet,ray2020diffusion,singh2020random,ahmad2020role,radice2021one,faisant2021optimal,abdoli2021stochastic,santra2020run,calvert2021searching,mercado2021search,bonomo2021first,tucci2022first,chen2022first,ahmad2022first,pal2023random,radice2023effects}. 
Summarizing the behaviour of the special cases, Reuveni strictly proved in Ref. \cite{Reuveni_PRL_2016} that any random process for which there exists a non-zero optimal restart rate obeys the universal relation
\begin{equation}
\label{relation1}
    \frac{\sigma(T_{r_{\ast}})}{\langle T_{r_{\ast}}\rangle}=1,
\end{equation}
where $\langle T_{r_{\ast}}\rangle$ and $\sigma(T_{r_{\ast}})$ are the average and standard deviation of the completion time of process $T_{r_\ast}$ in the presence of a restart with optimal rate $r_\ast$.

Further, in a recent publication \cite{starkov2023universal} it was demonstrated that optimally restarted random processes also obey the following universal inequality
\begin{equation}
\label{relation2}
 \frac{\langle T_{r_\ast}^3\rangle}{\langle T_{r_\ast}\rangle^3}\ge 6.
\end{equation}
From the results of the same work it follows that the average completion time of a random process in the presence of an optimal Poisson restart does not exceed a quarter of the geometric average completion time of the same process in the absence of a restart, i.e.
\begin{equation}
\label{relation3}
\langle T_{r_\ast}\rangle\ge\frac14    \langle T^{-1}\rangle^{-1}.
\end{equation}
Moreover, if the restart-free random completion time $T$ has a unimodal probability density $P(T)$, then the following inequality constraint is valid
\cite{starkov2023universal} 
\begin{eqnarray}
\label{relation4}
    \langle T_{r_\ast}\rangle\ge\frac14M,
\end{eqnarray}
where $M$ is a mode -- the value of the random variable $T$ that occurs most frequently.

Let us note in pass that the relations (\ref{relation1}) and (\ref{relation2}) follows from the extremum conditions $d \langle T_{r}\rangle/dr|_{r=r_\ast}=0$ and $d ^ 2 \langle T_{r}\rangle/dr^2|_{r=r_\ast}\ge 0$, respectively.
As for the equations (\ref{relation3}) and (\ref{relation4}), they are valid not only for the Poisson restart, but for any restart protocol.
Their proofs are based on the special properties of periodic restart, for more details see  \cite{starkov2023universal}.

Being insensitive to the process details, the relaitions (\ref{relation1}), (\ref{relation2}), (\ref{relation3}) and (\ref{relation4}) give model-independent predictions regarding the statistical characteristics of the random processes of different nature under optimally tuned Poisson restart.

In this work, we derive a number of novel universal relations that govern any random process at optimal restart conditions.
More specifically, we obtain a lower bound for the optimal rate, an upper bound for the resulting expected completion time, and a lower bound for the skewness and kurtosis coefficients of the distribution density of the random completion time.
In addition, we derive a universal inequality in the success probability optimization problem for a process with two alternative completion scenarios.
%Часть представленных результатов является прямым следствием уже известных раннее, в то время как для вывода других нам потребовалось провести дополнительные вычисления, детали которой можно найти в файле с дополнительной информацией.

%Model formulation. 
%%С рядом оговорок, влияние перезапуска на скорость протекания процессов самой разной природы может быть проанализировано в в рамках следующей  модели. 
%Рассмотрим случайный процесс, чье время завершения $T$ в отсутствии перезапуска характеризуется плотностью распределения $P(T)$.
%Под действием стохастического перезапуска с рейтом $r$ случайное время завершения процесса $T_r$ подчиняется следующему уравнению 
%\begin{equation}
%    T_r=TI(T\le R)+(R+T_r')I(T>R),
%\end{equation}
%где $R$ - экспоненциально распределенное с плотностью $\rho(R)=re^{-rR}$ случайная величина, задающая момент времени первого перезапуска; 
%$I(...)$ - индикаторная %функция, которая равна 1, если выполнено неравенство, стоящее в ее аргумента, и равна нулю в противном случае;
%$T_r'$ - статистически-%независимая реализация случайного времени $T_r$.

%Усредняя обе части уравнения по статистике случайных величин $T$ и $R$, а также используя равенства , получаем (см. детали в работе [??])
%\begin{equation}\label{eq:T_r_mean_main}
%\langle T_r\rangle=\frac{1-\tilde{P}(r)}{r\tilde{P}(r)},
%\end{equation}
%where $\tilde{P}(r)=\int_0^\infty dT P(T)e^{-rT}$ -- is the Laplace transform of the probability density $P(T)$. 

\textit{Lower bound on the optimal restart rate}.
If the distribution function $P(T)$ is known, then by minimizing the average completion time $\langle T_r\rangle$, given by the equation (\ref{eq:T_r_mean_main}), with respect to $r$, we can determine the  optimal rate $r_{\ast }$.
In a number of cases listed in the work \cite{Rotbart_2015}, such a problem allows an exact or approximate analytical solution.

Is it possible to estimate the optimal rate $r_\ast$ without knowing the detailed statistics of the restarted process?
Let us show that the lower bound on $r_\ast$ can be constructed based on information about the first three statistical moments $\langle T\rangle$, $\langle T^2\rangle$ and $\langle T^3\rangle$ (provided  they are finite).
Indeed, from (\ref{eq:T_r_mean_main}) we find
\begin{equation}
    \frac{d \langle T_r\rangle}{dr}=\frac{\tilde P^2(r)-\tilde P(r)-r\partial_r\tilde P(r)}{r^2\tilde P^2(r)}.
\end{equation}
Estimating the Laplace transform and its derivative using the inequalities $\tilde P(r)\le 1-r\langle T\rangle+\frac{r^2}{2}\langle T^2\rangle$ \cite{Zubkov_1999} and
$\partial_r \tilde P(r)\ge -\langle T\rangle+r\langle T^2\rangle-\frac{r^2}{2}\langle T^3\rangle$ \cite{starkov2023universal}, we can write
\begin{equation}
    \frac{d \langle T_r\rangle}{dr}\le \frac{(\frac{r}{2}\langle T^2\rangle-\langle T\rangle)^2+\frac{r}{2}\langle T^3\rangle-\frac{1}{2}\langle T^2\rangle}{r^2\tilde P^2(r)}.
\end{equation}
Next, by solving equation $(\frac{r}{2}\langle T^2\rangle-\langle T\rangle)^2+\frac{r}{2}\langle T^3\rangle-\frac{ 1}{2}\langle T^2\rangle\le0$ relative to $r$, we find that when $\sigma(T)>\langle T\rangle$ the average completion time $\langle T_r\rangle$ is guaranteed to decrease on the interval $0<r<\sqrt{\frac{\langle T^3\rangle^2}{\langle T^2\rangle^4}-\frac{4\langle T\rangle\langle T^3\rangle}{\langle T^2\rangle^3}+\frac{2}{\langle T^2\rangle}}-\frac{\langle T^3\rangle}{\langle T^2\rangle^2}+\frac{2\langle T\rangle}{\langle T^2\rangle}$.
%\begin{widetext}
%    \begin{equation}
%    0<r<\sqrt{\frac{\langle T^3\rangle^2}{\langle T^2\rangle^4}-\frac{4\langle T\rangle\langle T^3\rangle}{\langle T^2\rangle^3}+\frac{2}{\langle T^2\rangle}}-\frac{\langle T^3\rangle}{\langle T^2\rangle^2}+\frac{2\langle T\rangle}{\langle T^2\rangle}.
%\end{equation}
%Отсюда можно заключить, что 
%\begin{equation}
%    r_\ast\ge 
%    \sqrt{\frac{\langle T^3\rangle^2}{\langle T^2\rangle^4}-\frac{4\langle T\rangle\langle T^3\rangle}{\langle T^2\rangle^3}+\frac{2}{\langle T^2\rangle}}-\frac{\langle T^3\rangle}{\langle T^2\rangle^2}+\frac{2\langle T\rangle}{\langle T^2\rangle}.
%\end{equation}
%\end{widetext}
From this we can conclude that
\begin{eqnarray}
\label{result1}
    r_\ast\ge 
    \sqrt{\frac{\langle T^3\rangle^2}{\langle T^2\rangle^4}-\frac{4\langle T\rangle\langle T^3\rangle}{\langle T^2\rangle^3}+\frac{2}{\langle T^2\rangle}}-\frac{\langle T^3\rangle}{\langle T^2\rangle^2}+\frac{2\langle T\rangle}{\langle T^2\rangle}.
\end{eqnarray}

Let us discuss the scope of application of the obtained result.
The assumpton $\sigma(T)>\langle T\rangle$ used in the derivation of Eq.  (\ref{result1}) represents a sufficient (not necessary) condition for the effectiveness of Poisson restart \cite{Rotbart_2015, Reuveni_PRL_2016,Pal_JPA_2022}.
This assumption guarantees that the expression in the right side of Eq.  (\ref{result1}) is a real positive quantity.
Thus, (\ref{result1}) cannot be used for processes for which $\sigma(T)<\langle T\rangle$ but nevertheless there is a non-zero restart rate at which the global minimum of $\langle T_r\rangle$ is reached.
A simple example of such a process is given by the following completion time probability density $P(T)=0.5 e^{-rT}+0.5 e^{-10rT}$.
Next, the formula (\ref{result1}) is not applicable when at least one of   the statistical moments $\langle T\rangle$, $\langle T\rangle$ and $\langle T^3\rangle$  diverges.
This is exactly the case for distributions with fairly heavy tails of completion time distribution, which is typical for many probabilistic algorithms and random search processes.
Despite the noted shortcomings, inequality(\ref{result1}), to our knowledge, is the first rigorous result that allows to bound from below the optimal restart rate based on partial information about the statistics of original process.

Is it possible, in addition to the lower bound (\ref{result1}), to construct an upper estimate of the form $r_\ast\le r_{up}$, where $r_{up}$ is expressed in terms of the first three moments $\langle T\rangle$, $\langle T\rangle$ and $\langle T^3\rangle$?
The answer is no since for fixed values of these moments the optimal rate $r_\ast$ can be arbitrarily large, as can be seen in the following example.
Consider, probability density $P(T)=p\delta(T-t_1)+(1-p)\delta(T-t_2)$, where  $t_2= \frac{\mu^2+ \sigma^2}{\mu}$, $p=\frac{\sigma^2}{\mu^2 +\sigma^2}$, and  $t_1\to 0$.
It is easy to check that $\langle T\rangle=\mu$, $\langle T^2\rangle=\sigma^2+\mu^2$ and $\langle T^3\rangle=\frac{(\mu^2+\sigma^2)^2}{\mu}$.
As follows from (\ref{eq:T_r_mean_main}), if  $\sigma>\mu$, then there exists optimal $r_\ast\sim t^{-1}_1\to+\infty$ bringing $\langle T_{r}\rangle$ to minimum $\langle T_{r_\ast}\rangle$. 
%Since 
%$\langle T_r\rangle\to 0$ for 
%$r=t_1^{-1}$ and $\langle %T_r\rangle\le \langle %T_{r_\ast}\rangle$ we imediatelly conclude that 
% $\langle T_r\rangle\to 0$ at %$t_1\to0$.
%This argument proves that for the fixed  values of $\mu$ and $\sigma$, the optimal rate  $r_\ast\sim t^{-1}$ can be arbitrarily large.
Therefore,  troika  $(\langle T\rangle$,$\langle T^2\rangle$,$\langle T^3\rangle)$ does not produce any non trivial  upper bound for $r_\ast$.

%Исследуем, насколько близки реальные значения $r_\ast$ к построенной нижней границе для конкретных классов распределений.

%$\frac{\sqrt{\langle T^3\rangle^2-4\langle T\rangle\langle T^2\rangle\langle T^3\rangle+2\langle T^2\rangle^3}-\langle T^3\rangle+2\langle T\rangle \langle T^2\rangle}{\langle T^2\rangle^2}$

\textit{Bounds on average completion time under optimal restart rate}. 
In addition to the previously known inequality (\ref{relation3}) specifying the lower limit on $\langle T_{r_\ast}\rangle$, it is also possible to determine the upper bound limit.
%вопрос о том, насколько сильно оптимальный стохастический перезапуск понижает среднее время завершения в сравнении с его невозмущенным значением $\langle T\rangle$. 
As was shown in the work \cite{nikitin2023choosing}, if the condition $\sigma(T)>\langle T\rangle$ is met, then restart with the rate $r_0=\frac{1}{\langle T\rangle}-\sqrt{\frac{1}{\langle T\rangle^2}-\frac{3\langle T^2\rangle}{\langle T\rangle\langle T^3\rangle}+\frac{6\langle T\rangle}{\langle T^3\rangle}}$
%\begin{equation}
%    r_0=\frac{\langle T^3\rangle-\sqrt{\langle T^3\rangle(\langle T^3\rangle-3\langle T^2\rangle\langle T\rangle+6\langle T\rangle^3)}}{\langle T\rangle\langle T^3\rangle}
%\end{equation}
%\begin{equation}
%    r_0=\frac{1}{\langle T\rangle}-\sqrt{\frac{1}{\langle T\rangle^2}-\frac{3\langle T^2\rangle}{\langle T\rangle\langle T^3\rangle}+\frac{6\langle T\rangle}{\langle T^3\rangle}}
%\end{equation}
is guaranteed to reduce the average process completion time, and $\langle T_{r_0}\rangle\le \frac{3\langle T^2\rangle-2\langle T^3\rangle r_0}{6\langle T\rangle}$ .
By definition of the optimal rate $\langle T_{r_\ast}\rangle\le \langle T_{r}\rangle$ for any $r\ge0$, which means
\begin{equation}
\label{result2}
\langle T_{r_\ast}\rangle\le \frac{\langle T^2\rangle}{2\langle T\rangle}-\frac{\langle T^3\rangle}{3\langle T\rangle^2}+\frac13\sqrt{\frac{\langle T^3\rangle^2}{\langle T\rangle^4}-\frac{3\langle T^2\rangle \langle T^3\rangle}{\langle T\rangle^3}+\frac{6\langle T^3\rangle}{\langle T\rangle}}.
\end{equation}
%\begin{widetext}
%\begin{equation}
%\langle T_{r_\ast}\rangle\le \frac{3\langle T\rangle\langle T^2\rangle-2\langle T^3\rangle+2\sqrt{\langle T^3\rangle(\langle T^3\rangle-3\langle T^2\rangle\langle T\rangle+6\langle T\rangle^3)}}{6\langle T\rangle^2}.
%\end{equation}
%\end{widetext}
This is the upper bound announced just above for the mean completion time in the presence of an optimally tuned Poisson restart.
We emphasize that the expression in the right side of (\ref{result2}) does not exceed $\langle T\rangle$ if the sufficient condition for the efficiency of  restart $\sigma(T)>\langle T\rangle$ is satisfied.

Like the formula (\ref{result1}), the relation (\ref{result2}) becomes uninformative in situations where one or more of the first three statistical moments do not exist.
In this regard, it is of interest to construct an upper bound on $\langle T_{r_\ast}\rangle$, which would be well defined for any process.
Let us show that such a boundary can be expressed in terms of the median completion time of the unperturbed process $m$, defined by the condition $\int_0^{m}P(T)dT=1/2$.

Since $1-\tilde P(r)<1$ and $\tilde P(r)\ge \int_0^{r^{-1}}P(T)e^{-rT}dT\ge e^{-1} \int_0^{r^{-1}}P(T)dT$, then taking into account Eq.  (\ref{eq:T_r_mean_main}) we obtain
$\langle T_r\rangle <\frac{e}{r\int_0^{r^{-1}}P(T)dT}$.
Substituting $r=m^{-1}$ and recalling that $\langle T_{r_\ast}\rangle\le\langle T_r\rangle$ for any  $r\ge 0$, we finally arrive at the inequality 
\begin{eqnarray}
\label{result3}
    \langle T_{r_\ast}\rangle\le 2e m.
\end{eqnarray}
Note that in the work \cite{starkov2023universal} a similar result was obtained for periodic  restart: $ \langle T_{\tau_\ast}\rangle\le 2m$, where $T_{\tau}$ is the random completion time of the process being restarted every $\tau$ units of time,  and $\tau_\ast\ge 0$ is a restart period that minimizes the mean completion time.
%Разница в численным префакторах находится в согласии с тем фактов, что оптимальный периодический перезапуск является самой эффективным среди всех 

The inequality (\ref{result3}) allows us to formulate a new sufficient condition for the efficiency of Poisson restart.
More specifically, if
\begin{eqnarray}
\label{suff_condition}
\langle T\rangle> 2e m,    
\end{eqnarray}
then zero rate $r=0$ cannot be optimal, and, therefore, there should exist a value $r_{\ast}>0$, which reduces the average completion time of the process.
It is easy to see that this criterion is not equivalent to the aforementioned sufficient condition $\sigma(T)>\langle T\rangle$.
Indeed, if $P(T)=0.6 e^{-rT}+0.4 e^{-10rT}$, then $\langle T\rangle> 2e m$, whereas $\sigma(T)<\langle T\rangle$.

Note also that
we can easily turn the sufficient condition (\ref{suff_condition}) into constructive criteria of restart efficiency:
when inequality $\langle T\rangle> 2e m$ is satisfied, restart rate belonging to the interval $2e\langle T\rangle^{-1}\le r\le m^{-1}$ leads to reduction of the mean completion time.

\textit{Lower estimates for the skewness and kurtosis coefficients of the random time distribution density $T_{r_\ast}$.} 
Previous studies teach us that optimally tuned stochastic restart renders very different processes somewhat similar to each other \cite{Reuveni_PRL_2016}.
Namely, as noted above, at the optimal restart rate $r_\ast$ 
the relative fluctuation of the completion time is equal to unity for any stochastic process, see Eq. (\ref{relation1}).
We may recall that unit relative fluctuation is the property  exponentially distributed random variables.
This observation, coupled with the fact that Poisson restart turns the power-law tails of the completion time probability density  into exponential tails \cite{evans2011diffusion}, may suggest that for any random process under the optimally tuned restart the statistics of the completion time  is exactly exponential.
This conclusion, however, is refuted by counterexamples, see \cite{Reuveni_PRL_2016}.

Although the complete statistics of the random variable $T_{r_\ast}$ is not universal, its integral characteristics obey some well-defined restrictions.
Let us define a two-dimensional configuration space, where each statistical distribution law is characterized by two parameters: the skewness and the kurtosis coefficients, which are defined as standardized moments
\begin{eqnarray}
\label{asymmetry_def}
     A_r= \frac{\langle (T_r- \langle T_r\rangle)^3\rangle}{\sigma^3(T_r)},
\end{eqnarray}
and
\begin{equation}
\label{kurtozis_def}
    K_r=\frac{\langle (T_r- \langle T_r\rangle)^4\rangle}{\sigma^4(T_r)},
\end{equation}
respectively.
The first of these metrics plays the role of the asymmetry degree of the unimodal probability density relatively to its average value.
The second metric measures the extremality of the deviations of a random variable from its average value.
In descriptive statistics, skewness and kurtosis  are often used to characterize the possible deviations of statistical data from normality.
%Для (полу)гауссова распределения справедливо $A\approx  1$ и $K_{norm}\approx3.87$.
In our case, it is more interesting to compare with the case of exponential statistics, for which $A_{exp}=2$ and $K_{exp}= 9$.
Note also that for a half-normal distribution one has $A\approx 1$ and $K_{norm}\approx3.87$, while in the general case we obtain $-\infty<A<\infty$, $K\ge A^2 +1$ \cite{pearson1916ix,sharma2015skewness}.

Let us demonstrate that relations (\ref{relation1}) and (\ref{relation2}) impose lower bound on $A_{r_\ast}$ and $K_{r_\ast}$.
Indeed, from (\ref{relation1}), (\ref{relation2})  and (\ref{asymmetry_def}) one finds
\begin{equation}
\label{result4}
   A_{r_{\ast}}=\frac{\langle T_{r_\ast}^3 \rangle -3\langle T_{r_\ast}\rangle \sigma^2(T_{r_\ast})-\langle T_{r_\ast}\rangle^3}{\sigma^3(T_{r_\ast})}\ge 2.
\end{equation}
Next, at $r=r_\ast$ 
due to inequality $K_r\ge A_r^2+1$  and Eq. (\ref{result4})
 we obtain 
\begin{equation}
\label{result5}
    K_{r_\ast}\ge 5.
\end{equation}

\textit{Success probability optimization}. 
Beyond the mean completion time optimization, stochastic restart can be also used to manipulate the probability of getting the desired outcome of a random process that has two alternative completion scenarios -- success and failure \cite{Belan_PRL_2018,pal2019first,singh2023escape,chechkin2018random,singh2023bernoulli}.
%Потенциал перезапуска не ограничивается возможностью контроля над скоростью  протекания случайных процессов.
%Так, в работе \cite{Belan_PRL_2018} было показано, что перезапуск также может использоваться в целях оптимизации вероятности желаемого исхода  случайного процесса, имеющего два альтернативных сценария завершения (см. также \cite{pal2019first,singh2023escape,chechkin2018random,singh2023bernoulli}).
For the optimally tuned restart rate  $r_\ast$, which minimizes the success probability $p_{r}$, the following universal identity  is valid (see \cite{Belan_PRL_2018})
\begin{equation}
\label{relation5}
    \langle T^s_{r_\ast}\rangle=\langle T^f_{r_\ast}\rangle,
\end{equation}
where $\langle T^s_{r_\ast}\rangle$  and $\langle T^f_{r_\ast}\rangle$   denote the average completion time of, respectively,  successful and unsuccessful process realizations.
Below we show that in addition to Eq. (\ref{relation5}), the following inequality for the quadratic statistical moments of completion times is also valid at the optimal restart rate
\begin{equation}
\label{result6}
   \langle {T^s_{r_\ast}}^2\rangle\le  \langle {T^f_{r_\ast}}^2\rangle.
\end{equation}
%Кроме того, мы получим оценку снизу на оптимальный рейт $r_\ast$ и оценку сверху на оптимальную вероятность успеха $p_{r_\ast}$.

Let us derive Eq. (\ref{result6}).
The completion time probability density of a random process with two competing outcomes can be represented as $P(T)=P^s(T)+P^f(T)$, where $P^s(T)$  and $P^f(T)$ correspond the contributions of successful and unsuccessful trials, respectively.
The restart-free probability of success is then expressed as $p=\int_0^\infty P^s(T)dT$, and the mean time for successful runs is given by $\langle T^s\rangle=p^{-1}\int_0^\infty P^s(T)TdT$.
%Введем бинарную случайную величину, несущую информацию о результате протекания процесса: - если процесс завершился успехом  и в случае неудачи.
%В присутствии стохастического перезапуска эта случайная величина удовлетворяет уравнению 
%\begin{equation}
%    x_r=
%\end{equation}
%где ??
%Усредняя уравнение по статистике исходного случайного процесса и моментов перезапуска, находим вероятность успеха 
In the presence of stochastic restart with rate $r$, the probability of success becomes 
\begin{equation}
\label{p_r_exp}
p_r=\frac{\tilde P^s(r)}{\tilde P(r)}.
\end{equation}
Here $\tilde P^s(r)$ and $\tilde P(r)$ denote the Laplace-transforms of $P^s(T)$ and $P(T)$ evaluated at $r$.
The relation (\ref{relation5}) follows from Eq.(\ref{p_r_exp}) and extremality condition $d p_{r}/dr|_{r=r_\ast}=0$.

Now assume that a process which is already restarted at optimal rate $r_\ast$ becomes additionally subject to differentially small retart rate  $\delta r$.
Due to the additivity of Poisson flows and the condition $d p_{r}/dr|_{r=r_\ast}=0$, we can write
\begin{equation}
\label{eq1}
    p_{r_\ast+\delta r}=p_{r_\ast}+\frac{\delta r^2}{2} \frac{d^2p_r}{dr^2}|_{r_\ast}+o(\delta r^2).
\end{equation}
At the same time, from (\ref{p_r_exp}) we find
\begin{eqnarray}
 &&p_{r_\ast+\delta r}=\frac{\tilde P^s_{r_\ast}(\delta r)}{\tilde P_{r_\ast}(\delta r)}=\frac{\int_0^\infty dT e^{-\delta rT}P_{r_\ast}^s(T)}{\int_0^\infty dT e^{-\delta rT}P_{r_\ast}(T)}  =\\
&&=\frac{1-\delta r\langle T^s_{r_\ast}\rangle +\frac{\delta r^2}{2}\langle T^{s2}_{r_\ast}\rangle+o(\delta r^2)}{1-\delta r\langle T_{r_\ast}\rangle +\frac{\delta r^2}{2}\langle T^{2}_{r_\ast}\rangle+o(\delta r^2)}p=\\
\label{eq2}
&&=p_{r_\ast}+\frac{\delta r^2}{2} p_{r_\ast}(1-p_{r_\ast})(\langle {T_{r_\ast}^{s}}^2\rangle-\langle {T_{r_\ast}^f}^{2}\rangle)+o(\delta r^2),
\end{eqnarray}
where $\tilde P_{r_\ast}(r)$ and $\tilde P^s_{r_\ast}(r)$ denote the Laplace transforms of completion time probability densities $P_{r_\ast}(T)$ and $ P^s_{r_\ast}(T)$ in the presence of restart rate $r_\ast$ and  we used an obvious identity  $\langle T^2_r\rangle=p\langle {T_r^s}^2\rangle+(1-p)\langle {T_r^f}^2\rangle$.
%где $P_{r_\ast}(T)$ и $P_{r_\ast}^s(T)$ -  функции плотности распределения  времени завершения процесса  $T_{r_\ast}$ и его удачных реализаций  $T_{r_\ast}^s$ в присутствии перезапуска с рейтом $r_\ast$. 
Comparing (\ref{eq1}) and (\ref{eq2}), we find $d^2 p_{r}/dr^2|_{r=r_\ast}=p_{r_\ast}(1-p_{r_\ast})(\langle {T_{r_\ast}^{s}}^2\rangle-\langle {T_{r_\ast}^f}^{2}\rangle)/2$.
The point of maximum obyes the condition $d^2 p_{r}/dr^2|_{r=r_\ast}\le0$, which immediately yields  Eq. (\ref{result6}). 
Thus, under conditions of optimal stochastic restart, the completion time of successful trials is characterized by smaller fluctuations compared to unsuccessful attempts.

If the experiment with two alternative outcomes is repeated over and over again, then one should optimize the rate of obtaining desirable outcomes $\nu_r$ rather than the probability of success $p_r$ in an individual trial. The former metric is  given by the relation
\begin{eqnarray}
    \nu_r=\frac{N^s_r(t)}{t},
\end{eqnarray}
where $N^s_r(t)$  is the number of successfull outcomes observed during sufficiently large time interval $t\gg\langle T\rangle$.
Due to the law o large numbers, $N_r^s=p_rN_r$ and $t=N_r(t)\langle T\rangle$, where $N(t)$ - is the total number of process completions during the interval $t$, so that
\begin{eqnarray}
\label{nu_r}
    \nu_r=\frac{p_r}{\langle T\rangle}.
\end{eqnarray}
Next, substituting relations  (\ref{eq:T_r_mean_main}) and (\ref{p_r_exp}) into (\ref{nu_r}) one gets 
\begin{eqnarray}
\nu_r=\frac{r\tilde P^s(r)}{1-\tilde P^s(r)}.
\end{eqnarray}
If $r_\ast$ is the optimal restart rate bringing $\nu_r$ to its maximum value, then
\begin{eqnarray}
    &&\nu_{r_\ast+\delta r}=\frac{\delta r\tilde P^s_{r_\ast}(\delta r)}{1-\tilde P^s_{r_\ast}(\delta r)}=\\
    &&=\frac{1-\delta r\langle T^s_{r_\ast}\rangle +\frac{\delta r^2}{2}\langle T^{s2}_{r_\ast}\rangle+o(\delta r^2)}{1 -\frac{\delta r}{2}\frac{\langle T^{2}_{r_\ast}\rangle}{\langle T_{r_\ast}\rangle}+\frac{\delta r^2}{6}\frac{\langle T_{r_\ast}^3\rangle}{\langle T_{r_\ast}\rangle}+o(\delta r^2)}\frac{p_{r_\ast}}{\langle T_{r_\ast}\rangle}=\\
    &&=\left[1+\left(\frac{1}{2}\frac{\langle T_r^2\rangle}{\langle T_r\rangle}-\langle T_r^s\rangle\right)\delta r+\right.\\
    &&+ \left(\frac{1}{4}\frac{\langle T_r^2\rangle^2}{\langle T_r\rangle^2}-\frac{1}{2}\frac{\langle T^2_r\rangle\langle T_r^s\rangle}{\langle T_r\rangle}+\frac{1}{2}\langle T_r^{s2}\rangle-\right.\\
    &&\left.\left. - \frac{1}{6}\frac{\langle T^3_r\rangle}{\langle T_r\rangle}\right)\delta r^2+o(\delta r^2)\right]\frac{p_{r_\ast}}{\langle T_{r_\ast}\rangle}.
\end{eqnarray}
Imposing the conditions $d\nu_{r}/dr|_{r=r_\ast}=0$ and $d^2\nu_{r}/dr^2|_{r=r_\ast}\le 0$ we readily find the identity
\begin{eqnarray}
    \langle T_{r_\ast}^s\rangle=\frac{1}{2}\frac{\langle T_{r_\ast}^{2}\rangle}{\langle T_{r_\ast}\rangle},
\end{eqnarray}
and inequality
\begin{eqnarray}
    \langle T_{r_\ast}^s\rangle^2\le \frac{1}{3}\frac{\langle T_{r_\ast}^{3}\rangle}{\langle T_{r_\ast}\rangle}.
\end{eqnarray}

%Теперь получим для оптимального рейта в задаче оптимизации вероятности успеха оценку на оптимальный рейт.
%Для этого, воспользовавшись неравенствами и уравнением , запишем 
%\begin{eqnarray}
%    \frac{dp_r}{dr}=...\ge
%\end{eqnarray}

%Подчеркнем, что данная оценка справедлива только при условии $\langle T^s\rangle >\langle T\rangle$.

%Наконец, покажем, как оценить эффективность оптимального пуассоновского перезапуска.
%В работе было показано, что при условии, перезапуск с рейтом $r_0=??$ гарантировано повышает шансы на успех, и при этом $p_{r_0}\le $.
%Так как $p_{r_\ast}\ge p_r$ для любого $r\ge 0$, 

\textit{Conclusion.}
Our analysis contributes to the understanding of the universal features inherent to random processes subject to optimal Poisson restart.
The main results of the work are given by the relations (\ref{result1}), (\ref{result2}), (\ref{result3}), (\ref{result4}), (\ref{result5}) and (\ref{result6}).
As a useful consequence of these results, we have obtained a new criterion for the efficiency of Poisson restart, see Eq. (\ref{suff_condition}).
Future research should answer the question of whether the limits set by our inequalities are achieved.
In addition, a practically important question is how the presented results will change when taking into account the non-zero time required to perform the restart.

\acknowledgments

The work was supported by the Russian Science Foundation,
project no. 22-72-10052.

%Графики:

%0) График зависимости $T_r$ от $r$ для нескольких случайных процессов, демонстрирующий насколько правее диктуемой формулой (8) грани лежит абсцисса точка минимума, а также насколько ниже границы (9) ордината этой точки.
%Выбрать несколько процессов с конечными первыми тремя моментами: Mix of two exponential distributions, Mix of three exponential distributions, Mix of two Erlang distributions, Sum of two delta functions, suffficiently fast decaying Pareto distribution, 

%1) Аналог диаграммы 1а из статьи Старкова. Можно добавить обобщенное распределение Леви-Смирнова.   

%2) Диаграмма в плоскости параметров $(A_{r_\ast},K_{r_\ast})$ для тех же тех распределений, что использовались на предыдущей диаграмме. Все точки должны попасть в область $K_r\ge A_r^2+1$, $K_{r_\ast}\ge 5$, $A_{r_{\ast}}\ge 2$.

\bibliography{Main_eng-2}

\begin{thebibliography}{53}%
\makeatletter
\providecommand \@ifxundefined [1]{%
 \@ifx{#1\undefined}
}%
\providecommand \@ifnum [1]{%
 \ifnum #1\expandafter \@firstoftwo
 \else \expandafter \@secondoftwo
 \fi
}%
\providecommand \@ifx [1]{%
 \ifx #1\expandafter \@firstoftwo
 \else \expandafter \@secondoftwo
 \fi
}%
\providecommand \natexlab [1]{#1}%
\providecommand \enquote  [1]{``#1''}%
\providecommand \bibnamefont  [1]{#1}%
\providecommand \bibfnamefont [1]{#1}%
\providecommand \citenamefont [1]{#1}%
\providecommand \href@noop [0]{\@secondoftwo}%
\providecommand \href [0]{\begingroup \@sanitize@url \@href}%
\providecommand \@href[1]{\@@startlink{#1}\@@href}%
\providecommand \@@href[1]{\endgroup#1\@@endlink}%
\providecommand \@sanitize@url [0]{\catcode `\\12\catcode `\$12\catcode
  `\&12\catcode `\#12\catcode `\^12\catcode `\_12\catcode `\%12\relax}%
\providecommand \@@startlink[1]{}%
\providecommand \@@endlink[0]{}%
\providecommand \url  [0]{\begingroup\@sanitize@url \@url }%
\providecommand \@url [1]{\endgroup\@href {#1}{\urlprefix }}%
\providecommand \urlprefix  [0]{URL }%
\providecommand \Eprint [0]{\href }%
\providecommand \doibase [0]{http://dx.doi.org/}%
\providecommand \selectlanguage [0]{\@gobble}%
\providecommand \bibinfo  [0]{\@secondoftwo}%
\providecommand \bibfield  [0]{\@secondoftwo}%
\providecommand \translation [1]{[#1]}%
\providecommand \BibitemOpen [0]{}%
\providecommand \bibitemStop [0]{}%
\providecommand \bibitemNoStop [0]{.\EOS\space}%
\providecommand \EOS [0]{\spacefactor3000\relax}%
\providecommand \BibitemShut  [1]{\csname bibitem#1\endcsname}%
\let\auto@bib@innerbib\@empty
%</preamble>
\bibitem [{\citenamefont {Alt}\ \emph {et~al.}(1991)\citenamefont {Alt},
  \citenamefont {Guibas}, \citenamefont {Mehlhorn}, \citenamefont {Karp},\ and\
  \citenamefont {Wigderson}}]{Alt_1991}%
  \BibitemOpen
  \bibfield  {author} {\bibinfo {author} {\bibfnamefont {H.}~\bibnamefont
  {Alt}}, \bibinfo {author} {\bibfnamefont {L.}~\bibnamefont {Guibas}},
  \bibinfo {author} {\bibfnamefont {K.}~\bibnamefont {Mehlhorn}}, \bibinfo
  {author} {\bibfnamefont {R.}~\bibnamefont {Karp}}, \ and\ \bibinfo {author}
  {\bibfnamefont {A.}~\bibnamefont {Wigderson}},\ }\href@noop {} {\bibfield
  {journal} {\bibinfo  {journal} {Technical Report TR-91-057}\ } (\bibinfo
  {year} {1991})}\BibitemShut {NoStop}%
\bibitem [{\citenamefont {Luby}\ \emph {et~al.}(1993)\citenamefont {Luby},
  \citenamefont {Sinclair},\ and\ \citenamefont {Zuckerman}}]{Luby_1993}%
  \BibitemOpen
  \bibfield  {author} {\bibinfo {author} {\bibfnamefont {M.}~\bibnamefont
  {Luby}}, \bibinfo {author} {\bibfnamefont {A.}~\bibnamefont {Sinclair}}, \
  and\ \bibinfo {author} {\bibfnamefont {D.}~\bibnamefont {Zuckerman}},\
  }\href@noop {} {\bibfield  {journal} {\bibinfo  {journal} {Information
  Processing Letters}\ }\textbf {\bibinfo {volume} {47}},\ \bibinfo {pages}
  {173} (\bibinfo {year} {1993})}\BibitemShut {NoStop}%
\bibitem [{\citenamefont {Reuveni}\ \emph {et~al.}(2014)\citenamefont
  {Reuveni}, \citenamefont {Urbakh},\ and\ \citenamefont
  {Klafter}}]{Reuveni_2014}%
  \BibitemOpen
  \bibfield  {author} {\bibinfo {author} {\bibfnamefont {S.}~\bibnamefont
  {Reuveni}}, \bibinfo {author} {\bibfnamefont {M.}~\bibnamefont {Urbakh}}, \
  and\ \bibinfo {author} {\bibfnamefont {J.}~\bibnamefont {Klafter}},\
  }\href@noop {} {\bibfield  {journal} {\bibinfo  {journal} {Proceedings of the
  National Academy of Sciences}\ }\textbf {\bibinfo {volume} {111}},\ \bibinfo
  {pages} {4391} (\bibinfo {year} {2014})}\BibitemShut {NoStop}%
\bibitem [{\citenamefont {Rotbart}\ \emph {et~al.}(2015)\citenamefont
  {Rotbart}, \citenamefont {Reuveni},\ and\ \citenamefont
  {Urbakh}}]{Rotbart_2015}%
  \BibitemOpen
  \bibfield  {author} {\bibinfo {author} {\bibfnamefont {T.}~\bibnamefont
  {Rotbart}}, \bibinfo {author} {\bibfnamefont {S.}~\bibnamefont {Reuveni}}, \
  and\ \bibinfo {author} {\bibfnamefont {M.}~\bibnamefont {Urbakh}},\
  }\href@noop {} {\bibfield  {journal} {\bibinfo  {journal} {Physical Review
  E}\ }\textbf {\bibinfo {volume} {92}},\ \bibinfo {pages} {060101} (\bibinfo
  {year} {2015})}\BibitemShut {NoStop}%
\bibitem [{\citenamefont {Reuveni}(2016)}]{Reuveni_PRL_2016}%
  \BibitemOpen
  \bibfield  {author} {\bibinfo {author} {\bibfnamefont {S.}~\bibnamefont
  {Reuveni}},\ }\href@noop {} {\bibfield  {journal} {\bibinfo  {journal}
  {Physical Review Letters}\ }\textbf {\bibinfo {volume} {116}},\ \bibinfo
  {pages} {170601} (\bibinfo {year} {2016})}\BibitemShut {NoStop}%
\bibitem [{\citenamefont {Berezhkovskii}\ \emph {et~al.}(2017)\citenamefont
  {Berezhkovskii}, \citenamefont {Dagdug},\ and\ \citenamefont
  {Bezrukov}}]{berezhkovskii2017bulk}%
  \BibitemOpen
  \bibfield  {author} {\bibinfo {author} {\bibfnamefont {A.~M.}\ \bibnamefont
  {Berezhkovskii}}, \bibinfo {author} {\bibfnamefont {L.}~\bibnamefont
  {Dagdug}}, \ and\ \bibinfo {author} {\bibfnamefont {S.~M.}\ \bibnamefont
  {Bezrukov}},\ }\href@noop {} {\bibfield  {journal} {\bibinfo  {journal} {The
  Journal of Chemical Physics}\ }\textbf {\bibinfo {volume} {147}} (\bibinfo
  {year} {2017})}\BibitemShut {NoStop}%
\bibitem [{\citenamefont {Wu}\ and\ \citenamefont {Beek}(2007)}]{Wu_2007}%
  \BibitemOpen
  \bibfield  {author} {\bibinfo {author} {\bibfnamefont {H.}~\bibnamefont
  {Wu}}\ and\ \bibinfo {author} {\bibfnamefont {P.}~\bibnamefont {Beek}},\
  }\href@noop {} {\bibfield  {journal} {\bibinfo  {journal} {In International
  Conference on Principles and Practice of Constraint Programming}\ ,\ \bibinfo
  {pages} {681}} (\bibinfo {year} {2007})}\BibitemShut {NoStop}%
\bibitem [{\citenamefont {Lorenz}(2018)}]{Lorenz_2018}%
  \BibitemOpen
  \bibfield  {author} {\bibinfo {author} {\bibfnamefont {J.}~\bibnamefont
  {Lorenz}},\ }\href@noop {} {\bibfield  {journal} {\bibinfo  {journal} {In
  International Conference on Current Trends in Theory and Practice of
  Informatics}\ ,\ \bibinfo {pages} {493}} (\bibinfo {year}
  {2018})}\BibitemShut {NoStop}%
\bibitem [{\citenamefont {Lorenz}(2021)}]{Lorenz_2021}%
  \BibitemOpen
  \bibfield  {author} {\bibinfo {author} {\bibfnamefont {J.}~\bibnamefont
  {Lorenz}},\ }\href@noop {} {\bibfield  {journal} {\bibinfo  {journal} {Theory
  of Computing Systems}\ }\textbf {\bibinfo {volume} {65}},\ \bibinfo {pages}
  {1143} (\bibinfo {year} {2021})}\BibitemShut {NoStop}%
\bibitem [{\citenamefont {Schulte}\ \emph {et~al.}(2010)\citenamefont
  {Schulte}, \citenamefont {Tack},\ and\ \citenamefont
  {Lagerkvist}}]{schulte2010modeling}%
  \BibitemOpen
  \bibfield  {author} {\bibinfo {author} {\bibfnamefont {C.}~\bibnamefont
  {Schulte}}, \bibinfo {author} {\bibfnamefont {G.}~\bibnamefont {Tack}}, \
  and\ \bibinfo {author} {\bibfnamefont {M.~Z.}\ \bibnamefont {Lagerkvist}},\
  }\href@noop {} {\bibfield  {journal} {\bibinfo  {journal} {Schulte, Christian
  and Tack, Guido and Lagerkvist, Mikael}\ }\textbf {\bibinfo {volume} {1}}
  (\bibinfo {year} {2010})}\BibitemShut {NoStop}%
\bibitem [{\citenamefont {Cire}\ \emph {et~al.}(2014)\citenamefont {Cire},
  \citenamefont {Kadioglu},\ and\ \citenamefont {Sellmann}}]{cire2014parallel}%
  \BibitemOpen
  \bibfield  {author} {\bibinfo {author} {\bibfnamefont {A.}~\bibnamefont
  {Cire}}, \bibinfo {author} {\bibfnamefont {S.}~\bibnamefont {Kadioglu}}, \
  and\ \bibinfo {author} {\bibfnamefont {M.}~\bibnamefont {Sellmann}},\ }in\
  \href@noop {} {\emph {\bibinfo {booktitle} {Proceedings of the AAAI
  Conference on Artificial Intelligence}}},\ Vol.~\bibinfo {volume} {28}\
  (\bibinfo {year} {2014})\BibitemShut {NoStop}%
\bibitem [{\citenamefont {Amadini}\ \emph {et~al.}(2018)\citenamefont
  {Amadini}, \citenamefont {Gabbrielli},\ and\ \citenamefont
  {Mauro}}]{amadini2018sunny}%
  \BibitemOpen
  \bibfield  {author} {\bibinfo {author} {\bibfnamefont {R.}~\bibnamefont
  {Amadini}}, \bibinfo {author} {\bibfnamefont {M.}~\bibnamefont {Gabbrielli}},
  \ and\ \bibinfo {author} {\bibfnamefont {J.}~\bibnamefont {Mauro}},\
  }\href@noop {} {\bibfield  {journal} {\bibinfo  {journal} {Theory and
  Practice of Logic Programming}\ }\textbf {\bibinfo {volume} {18}},\ \bibinfo
  {pages} {81} (\bibinfo {year} {2018})}\BibitemShut {NoStop}%
\bibitem [{\citenamefont {Van~Moorsel}\ and\ \citenamefont
  {Wolter}(2006)}]{van2006analysis}%
  \BibitemOpen
  \bibfield  {author} {\bibinfo {author} {\bibfnamefont {A.~P.}\ \bibnamefont
  {Van~Moorsel}}\ and\ \bibinfo {author} {\bibfnamefont {K.}~\bibnamefont
  {Wolter}},\ }\href@noop {} {\bibfield  {journal} {\bibinfo  {journal} {IEEE
  Transactions on Software Engineering}\ }\textbf {\bibinfo {volume} {32}},\
  \bibinfo {pages} {547} (\bibinfo {year} {2006})}\BibitemShut {NoStop}%
\bibitem [{\citenamefont {Wallace}\ and\ \citenamefont
  {Wallace}(2020)}]{wallace2020search}%
  \BibitemOpen
  \bibfield  {author} {\bibinfo {author} {\bibfnamefont {M.}~\bibnamefont
  {Wallace}}\ and\ \bibinfo {author} {\bibfnamefont {M.}~\bibnamefont
  {Wallace}},\ }\href@noop {} {\bibfield  {journal} {\bibinfo  {journal}
  {Building Decision Support Systems: using MiniZinc}\ ,\ \bibinfo {pages}
  {165}} (\bibinfo {year} {2020})}\BibitemShut {NoStop}%
\bibitem [{\citenamefont {Schroeder}\ and\ \citenamefont
  {Buro}(2001)}]{schroeder2001does}%
  \BibitemOpen
  \bibfield  {author} {\bibinfo {author} {\bibfnamefont {M.}~\bibnamefont
  {Schroeder}}\ and\ \bibinfo {author} {\bibfnamefont {L.}~\bibnamefont
  {Buro}},\ }in\ \href@noop {} {\emph {\bibinfo {booktitle} {Proceedings of the
  Workshop on Infrastructure for Agents, MAS, and Scalable MAS at the
  Conference Autonomous Agents}}}\ (\bibinfo {year} {2001})\BibitemShut
  {NoStop}%
\bibitem [{\citenamefont {Evans}\ and\ \citenamefont
  {Majumdar}(2011{\natexlab{a}})}]{EM_2011}%
  \BibitemOpen
  \bibfield  {author} {\bibinfo {author} {\bibfnamefont {M.}~\bibnamefont
  {Evans}}\ and\ \bibinfo {author} {\bibfnamefont {S.}~\bibnamefont
  {Majumdar}},\ }\href@noop {} {\bibfield  {journal} {\bibinfo  {journal}
  {Physical Review Letters}\ }\textbf {\bibinfo {volume} {106}},\ \bibinfo
  {pages} {160601} (\bibinfo {year} {2011}{\natexlab{a}})}\BibitemShut
  {NoStop}%
\bibitem [{\citenamefont {Evans}\ \emph {et~al.}(2020)\citenamefont {Evans},
  \citenamefont {Majumdar},\ and\ \citenamefont
  {Schehr}}]{evans2020stochastic}%
  \BibitemOpen
  \bibfield  {author} {\bibinfo {author} {\bibfnamefont {M.~R.}\ \bibnamefont
  {Evans}}, \bibinfo {author} {\bibfnamefont {S.~N.}\ \bibnamefont {Majumdar}},
  \ and\ \bibinfo {author} {\bibfnamefont {G.}~\bibnamefont {Schehr}},\
  }\href@noop {} {\bibfield  {journal} {\bibinfo  {journal} {Journal of Physics
  A: Mathematical and Theoretical}\ }\textbf {\bibinfo {volume} {53}},\
  \bibinfo {pages} {193001} (\bibinfo {year} {2020})}\BibitemShut {NoStop}%
\bibitem [{\citenamefont {Evans}\ and\ \citenamefont
  {Majumdar}(2011{\natexlab{b}})}]{evans2011diffusion}%
  \BibitemOpen
  \bibfield  {author} {\bibinfo {author} {\bibfnamefont {M.~R.}\ \bibnamefont
  {Evans}}\ and\ \bibinfo {author} {\bibfnamefont {S.~N.}\ \bibnamefont
  {Majumdar}},\ }\href@noop {} {\bibfield  {journal} {\bibinfo  {journal}
  {Journal of Physics A: Mathematical and Theoretical}\ }\textbf {\bibinfo
  {volume} {44}},\ \bibinfo {pages} {435001} (\bibinfo {year}
  {2011}{\natexlab{b}})}\BibitemShut {NoStop}%
\bibitem [{\citenamefont {Evans}\ and\ \citenamefont
  {Majumdar}(2014)}]{evans2014diffusion}%
  \BibitemOpen
  \bibfield  {author} {\bibinfo {author} {\bibfnamefont {M.~R.}\ \bibnamefont
  {Evans}}\ and\ \bibinfo {author} {\bibfnamefont {S.~N.}\ \bibnamefont
  {Majumdar}},\ }\href@noop {} {\bibfield  {journal} {\bibinfo  {journal}
  {Journal of Physics A: Mathematical and Theoretical}\ }\textbf {\bibinfo
  {volume} {47}},\ \bibinfo {pages} {285001} (\bibinfo {year}
  {2014})}\BibitemShut {NoStop}%
\bibitem [{\citenamefont {Evans}\ and\ \citenamefont
  {Majumdar}(2018)}]{evans2018run}%
  \BibitemOpen
  \bibfield  {author} {\bibinfo {author} {\bibfnamefont {M.~R.}\ \bibnamefont
  {Evans}}\ and\ \bibinfo {author} {\bibfnamefont {S.~N.}\ \bibnamefont
  {Majumdar}},\ }\href@noop {} {\bibfield  {journal} {\bibinfo  {journal}
  {Journal of Physics A: Mathematical and Theoretical}\ }\textbf {\bibinfo
  {volume} {51}},\ \bibinfo {pages} {475003} (\bibinfo {year}
  {2018})}\BibitemShut {NoStop}%
\bibitem [{\citenamefont {Blumer}\ \emph {et~al.}(2024)\citenamefont {Blumer},
  \citenamefont {Reuveni},\ and\ \citenamefont
  {Hirshberg}}]{blumer2024combining}%
  \BibitemOpen
  \bibfield  {author} {\bibinfo {author} {\bibfnamefont {O.}~\bibnamefont
  {Blumer}}, \bibinfo {author} {\bibfnamefont {S.}~\bibnamefont {Reuveni}}, \
  and\ \bibinfo {author} {\bibfnamefont {B.}~\bibnamefont {Hirshberg}},\
  }\href@noop {} {\bibfield  {journal} {\bibinfo  {journal} {Nature
  Communications}\ }\textbf {\bibinfo {volume} {15}},\ \bibinfo {pages} {240}
  (\bibinfo {year} {2024})}\BibitemShut {NoStop}%
\bibitem [{\citenamefont {Juli{\'a}n-Salgado}\ \emph
  {et~al.}(2024)\citenamefont {Juli{\'a}n-Salgado}, \citenamefont {Dagdug},\
  and\ \citenamefont {Boyer}}]{julian2024diffusion}%
  \BibitemOpen
  \bibfield  {author} {\bibinfo {author} {\bibfnamefont {P.}~\bibnamefont
  {Juli{\'a}n-Salgado}}, \bibinfo {author} {\bibfnamefont {L.}~\bibnamefont
  {Dagdug}}, \ and\ \bibinfo {author} {\bibfnamefont {D.}~\bibnamefont
  {Boyer}},\ }\href@noop {} {\bibfield  {journal} {\bibinfo  {journal}
  {Physical Review E}\ }\textbf {\bibinfo {volume} {109}},\ \bibinfo {pages}
  {024134} (\bibinfo {year} {2024})}\BibitemShut {NoStop}%
\bibitem [{\citenamefont {Pal}\ \emph {et~al.}(2022)\citenamefont {Pal},
  \citenamefont {Kostinski},\ and\ \citenamefont {Reuveni}}]{Pal_JPA_2022}%
  \BibitemOpen
  \bibfield  {author} {\bibinfo {author} {\bibfnamefont {A.}~\bibnamefont
  {Pal}}, \bibinfo {author} {\bibfnamefont {S.}~\bibnamefont {Kostinski}}, \
  and\ \bibinfo {author} {\bibfnamefont {S.}~\bibnamefont {Reuveni}},\
  }\href@noop {} {\bibfield  {journal} {\bibinfo  {journal} {Journal of Physics
  A: Mathematical and Theoretical}\ }\textbf {\bibinfo {volume} {55}},\
  \bibinfo {pages} {021001} (\bibinfo {year} {2022})}\BibitemShut {NoStop}%
\bibitem [{\citenamefont {Pal}\ and\ \citenamefont
  {Prasad}(2019{\natexlab{a}})}]{Pal_2019a}%
  \BibitemOpen
  \bibfield  {author} {\bibinfo {author} {\bibfnamefont {A.}~\bibnamefont
  {Pal}}\ and\ \bibinfo {author} {\bibfnamefont {V.}~\bibnamefont {Prasad}},\
  }\href@noop {} {\bibfield  {journal} {\bibinfo  {journal} {Physical Review
  E}\ }\textbf {\bibinfo {volume} {99}},\ \bibinfo {pages} {032123} (\bibinfo
  {year} {2019}{\natexlab{a}})}\BibitemShut {NoStop}%
\bibitem [{\citenamefont {Bonomo}\ and\ \citenamefont
  {Pal}(2021)}]{bonomo2021first}%
  \BibitemOpen
  \bibfield  {author} {\bibinfo {author} {\bibfnamefont {O.~L.}\ \bibnamefont
  {Bonomo}}\ and\ \bibinfo {author} {\bibfnamefont {A.}~\bibnamefont {Pal}},\
  }\href@noop {} {\bibfield  {journal} {\bibinfo  {journal} {Physical Review
  E}\ }\textbf {\bibinfo {volume} {103}},\ \bibinfo {pages} {052129} (\bibinfo
  {year} {2021})}\BibitemShut {NoStop}%
\bibitem [{\citenamefont {Ray}\ and\ \citenamefont
  {Reuveni}(2020)}]{ray2020diffusion}%
  \BibitemOpen
  \bibfield  {author} {\bibinfo {author} {\bibfnamefont {S.}~\bibnamefont
  {Ray}}\ and\ \bibinfo {author} {\bibfnamefont {S.}~\bibnamefont {Reuveni}},\
  }\href@noop {} {\bibfield  {journal} {\bibinfo  {journal} {The Journal of
  chemical physics}\ }\textbf {\bibinfo {volume} {152}} (\bibinfo {year}
  {2020})}\BibitemShut {NoStop}%
\bibitem [{\citenamefont {Ray}\ \emph {et~al.}(2019)\citenamefont {Ray},
  \citenamefont {Mondal},\ and\ \citenamefont {Reuveni}}]{ray2019peclet}%
  \BibitemOpen
  \bibfield  {author} {\bibinfo {author} {\bibfnamefont {S.}~\bibnamefont
  {Ray}}, \bibinfo {author} {\bibfnamefont {D.}~\bibnamefont {Mondal}}, \ and\
  \bibinfo {author} {\bibfnamefont {S.}~\bibnamefont {Reuveni}},\ }\href@noop
  {} {\bibfield  {journal} {\bibinfo  {journal} {Journal of Physics A:
  Mathematical and Theoretical}\ }\textbf {\bibinfo {volume} {52}},\ \bibinfo
  {pages} {255002} (\bibinfo {year} {2019})}\BibitemShut {NoStop}%
\bibitem [{\citenamefont {Whitehouse}\ \emph {et~al.}(2013)\citenamefont
  {Whitehouse}, \citenamefont {Evans},\ and\ \citenamefont
  {Majumdar}}]{whitehouse2013effect}%
  \BibitemOpen
  \bibfield  {author} {\bibinfo {author} {\bibfnamefont {J.}~\bibnamefont
  {Whitehouse}}, \bibinfo {author} {\bibfnamefont {M.~R.}\ \bibnamefont
  {Evans}}, \ and\ \bibinfo {author} {\bibfnamefont {S.~N.}\ \bibnamefont
  {Majumdar}},\ }\href@noop {} {\bibfield  {journal} {\bibinfo  {journal}
  {Physical Review E}\ }\textbf {\bibinfo {volume} {87}},\ \bibinfo {pages}
  {022118} (\bibinfo {year} {2013})}\BibitemShut {NoStop}%
\bibitem [{\citenamefont {Kusmierz}\ \emph {et~al.}(2014)\citenamefont
  {Kusmierz}, \citenamefont {Majumdar}, \citenamefont {Sabhapandit},\ and\
  \citenamefont {Schehr}}]{kusmierz2014first}%
  \BibitemOpen
  \bibfield  {author} {\bibinfo {author} {\bibfnamefont {L.}~\bibnamefont
  {Kusmierz}}, \bibinfo {author} {\bibfnamefont {S.~N.}\ \bibnamefont
  {Majumdar}}, \bibinfo {author} {\bibfnamefont {S.}~\bibnamefont
  {Sabhapandit}}, \ and\ \bibinfo {author} {\bibfnamefont {G.}~\bibnamefont
  {Schehr}},\ }\href@noop {} {\bibfield  {journal} {\bibinfo  {journal}
  {Physical review letters}\ }\textbf {\bibinfo {volume} {113}},\ \bibinfo
  {pages} {220602} (\bibinfo {year} {2014})}\BibitemShut {NoStop}%
\bibitem [{\citenamefont {Ku{\'s}mierz}\ and\ \citenamefont
  {Gudowska-Nowak}(2015)}]{kusmierz2015optimal}%
  \BibitemOpen
  \bibfield  {author} {\bibinfo {author} {\bibfnamefont {{\L}.}~\bibnamefont
  {Ku{\'s}mierz}}\ and\ \bibinfo {author} {\bibfnamefont {E.}~\bibnamefont
  {Gudowska-Nowak}},\ }\href@noop {} {\bibfield  {journal} {\bibinfo  {journal}
  {Physical Review E}\ }\textbf {\bibinfo {volume} {92}},\ \bibinfo {pages}
  {052127} (\bibinfo {year} {2015})}\BibitemShut {NoStop}%
\bibitem [{\citenamefont {Singh}(2020)}]{singh2020random}%
  \BibitemOpen
  \bibfield  {author} {\bibinfo {author} {\bibfnamefont {P.}~\bibnamefont
  {Singh}},\ }\href@noop {} {\bibfield  {journal} {\bibinfo  {journal} {Journal
  of Physics A: Mathematical and Theoretical}\ }\textbf {\bibinfo {volume}
  {53}},\ \bibinfo {pages} {405005} (\bibinfo {year} {2020})}\BibitemShut
  {NoStop}%
\bibitem [{\citenamefont {Ahmad}\ and\ \citenamefont
  {Das}(2020)}]{ahmad2020role}%
  \BibitemOpen
  \bibfield  {author} {\bibinfo {author} {\bibfnamefont {S.}~\bibnamefont
  {Ahmad}}\ and\ \bibinfo {author} {\bibfnamefont {D.}~\bibnamefont {Das}},\
  }\href@noop {} {\bibfield  {journal} {\bibinfo  {journal} {Physical Review
  E}\ }\textbf {\bibinfo {volume} {102}},\ \bibinfo {pages} {032145} (\bibinfo
  {year} {2020})}\BibitemShut {NoStop}%
\bibitem [{\citenamefont {Radice}(2021)}]{radice2021one}%
  \BibitemOpen
  \bibfield  {author} {\bibinfo {author} {\bibfnamefont {M.}~\bibnamefont
  {Radice}},\ }\href@noop {} {\bibfield  {journal} {\bibinfo  {journal}
  {Physical Review E}\ }\textbf {\bibinfo {volume} {104}},\ \bibinfo {pages}
  {044126} (\bibinfo {year} {2021})}\BibitemShut {NoStop}%
\bibitem [{\citenamefont {Faisant}\ \emph {et~al.}(2021)\citenamefont
  {Faisant}, \citenamefont {Besga}, \citenamefont {Petrosyan}, \citenamefont
  {Ciliberto},\ and\ \citenamefont {Majumdar}}]{faisant2021optimal}%
  \BibitemOpen
  \bibfield  {author} {\bibinfo {author} {\bibfnamefont {F.}~\bibnamefont
  {Faisant}}, \bibinfo {author} {\bibfnamefont {B.}~\bibnamefont {Besga}},
  \bibinfo {author} {\bibfnamefont {A.}~\bibnamefont {Petrosyan}}, \bibinfo
  {author} {\bibfnamefont {S.}~\bibnamefont {Ciliberto}}, \ and\ \bibinfo
  {author} {\bibfnamefont {S.~N.}\ \bibnamefont {Majumdar}},\ }\href@noop {}
  {\bibfield  {journal} {\bibinfo  {journal} {Journal of Statistical Mechanics:
  Theory and Experiment}\ }\textbf {\bibinfo {volume} {2021}},\ \bibinfo
  {pages} {113203} (\bibinfo {year} {2021})}\BibitemShut {NoStop}%
\bibitem [{\citenamefont {Abdoli}\ and\ \citenamefont
  {Sharma}(2021)}]{abdoli2021stochastic}%
  \BibitemOpen
  \bibfield  {author} {\bibinfo {author} {\bibfnamefont {I.}~\bibnamefont
  {Abdoli}}\ and\ \bibinfo {author} {\bibfnamefont {A.}~\bibnamefont
  {Sharma}},\ }\href@noop {} {\bibfield  {journal} {\bibinfo  {journal} {Soft
  Matter}\ }\textbf {\bibinfo {volume} {17}},\ \bibinfo {pages} {1307}
  (\bibinfo {year} {2021})}\BibitemShut {NoStop}%
\bibitem [{\citenamefont {Santra}\ \emph {et~al.}(2020)\citenamefont {Santra},
  \citenamefont {Basu},\ and\ \citenamefont {Sabhapandit}}]{santra2020run}%
  \BibitemOpen
  \bibfield  {author} {\bibinfo {author} {\bibfnamefont {I.}~\bibnamefont
  {Santra}}, \bibinfo {author} {\bibfnamefont {U.}~\bibnamefont {Basu}}, \ and\
  \bibinfo {author} {\bibfnamefont {S.}~\bibnamefont {Sabhapandit}},\
  }\href@noop {} {\bibfield  {journal} {\bibinfo  {journal} {Journal of
  Statistical Mechanics: Theory and Experiment}\ }\textbf {\bibinfo {volume}
  {2020}},\ \bibinfo {pages} {113206} (\bibinfo {year} {2020})}\BibitemShut
  {NoStop}%
\bibitem [{\citenamefont {Calvert}\ and\ \citenamefont
  {Evans}(2021)}]{calvert2021searching}%
  \BibitemOpen
  \bibfield  {author} {\bibinfo {author} {\bibfnamefont {G.~R.}\ \bibnamefont
  {Calvert}}\ and\ \bibinfo {author} {\bibfnamefont {M.~R.}\ \bibnamefont
  {Evans}},\ }\href@noop {} {\bibfield  {journal} {\bibinfo  {journal} {The
  European Physical Journal B}\ }\textbf {\bibinfo {volume} {94}},\ \bibinfo
  {pages} {1} (\bibinfo {year} {2021})}\BibitemShut {NoStop}%
\bibitem [{\citenamefont {Mercado-V{\'a}squez}\ and\ \citenamefont
  {Boyer}(2021)}]{mercado2021search}%
  \BibitemOpen
  \bibfield  {author} {\bibinfo {author} {\bibfnamefont {G.}~\bibnamefont
  {Mercado-V{\'a}squez}}\ and\ \bibinfo {author} {\bibfnamefont
  {D.}~\bibnamefont {Boyer}},\ }\href@noop {} {\bibfield  {journal} {\bibinfo
  {journal} {Journal of Physics A: Mathematical and Theoretical}\ }\textbf
  {\bibinfo {volume} {54}},\ \bibinfo {pages} {444002} (\bibinfo {year}
  {2021})}\BibitemShut {NoStop}%
\bibitem [{\citenamefont {Tucci}\ \emph {et~al.}(2022)\citenamefont {Tucci},
  \citenamefont {Gambassi}, \citenamefont {Majumdar},\ and\ \citenamefont
  {Schehr}}]{tucci2022first}%
  \BibitemOpen
  \bibfield  {author} {\bibinfo {author} {\bibfnamefont {G.}~\bibnamefont
  {Tucci}}, \bibinfo {author} {\bibfnamefont {A.}~\bibnamefont {Gambassi}},
  \bibinfo {author} {\bibfnamefont {S.~N.}\ \bibnamefont {Majumdar}}, \ and\
  \bibinfo {author} {\bibfnamefont {G.}~\bibnamefont {Schehr}},\ }\href@noop {}
  {\bibfield  {journal} {\bibinfo  {journal} {Physical Review E}\ }\textbf
  {\bibinfo {volume} {106}},\ \bibinfo {pages} {044127} (\bibinfo {year}
  {2022})}\BibitemShut {NoStop}%
\bibitem [{\citenamefont {Chen}\ and\ \citenamefont
  {Huang}(2022)}]{chen2022first}%
  \BibitemOpen
  \bibfield  {author} {\bibinfo {author} {\bibfnamefont {H.}~\bibnamefont
  {Chen}}\ and\ \bibinfo {author} {\bibfnamefont {F.}~\bibnamefont {Huang}},\
  }\href@noop {} {\bibfield  {journal} {\bibinfo  {journal} {Physical Review
  E}\ }\textbf {\bibinfo {volume} {105}},\ \bibinfo {pages} {034109} (\bibinfo
  {year} {2022})}\BibitemShut {NoStop}%
\bibitem [{\citenamefont {Ahmad}\ \emph {et~al.}(2022)\citenamefont {Ahmad},
  \citenamefont {Rijal},\ and\ \citenamefont {Das}}]{ahmad2022first}%
  \BibitemOpen
  \bibfield  {author} {\bibinfo {author} {\bibfnamefont {S.}~\bibnamefont
  {Ahmad}}, \bibinfo {author} {\bibfnamefont {K.}~\bibnamefont {Rijal}}, \ and\
  \bibinfo {author} {\bibfnamefont {D.}~\bibnamefont {Das}},\ }\href@noop {}
  {\bibfield  {journal} {\bibinfo  {journal} {Physical Review E}\ }\textbf
  {\bibinfo {volume} {105}},\ \bibinfo {pages} {044134} (\bibinfo {year}
  {2022})}\BibitemShut {NoStop}%
\bibitem [{\citenamefont {Pal}\ \emph {et~al.}(2023)\citenamefont {Pal},
  \citenamefont {Stojkoski},\ and\ \citenamefont {Sandev}}]{pal2023random}%
  \BibitemOpen
  \bibfield  {author} {\bibinfo {author} {\bibfnamefont {A.}~\bibnamefont
  {Pal}}, \bibinfo {author} {\bibfnamefont {V.}~\bibnamefont {Stojkoski}}, \
  and\ \bibinfo {author} {\bibfnamefont {T.}~\bibnamefont {Sandev}},\
  }\href@noop {} {\bibfield  {journal} {\bibinfo  {journal} {arXiv preprint
  arXiv:2310.12057}\ } (\bibinfo {year} {2023})}\BibitemShut {NoStop}%
\bibitem [{\citenamefont {Radice}(2023)}]{radice2023effects}%
  \BibitemOpen
  \bibfield  {author} {\bibinfo {author} {\bibfnamefont {M.}~\bibnamefont
  {Radice}},\ }\href@noop {} {\bibfield  {journal} {\bibinfo  {journal}
  {Physical Review E}\ }\textbf {\bibinfo {volume} {107}},\ \bibinfo {pages}
  {024136} (\bibinfo {year} {2023})}\BibitemShut {NoStop}%
\bibitem [{\citenamefont {Starkov}\ and\ \citenamefont
  {Belan}(2023)}]{starkov2023universal}%
  \BibitemOpen
  \bibfield  {author} {\bibinfo {author} {\bibfnamefont {D.}~\bibnamefont
  {Starkov}}\ and\ \bibinfo {author} {\bibfnamefont {S.}~\bibnamefont
  {Belan}},\ }\href@noop {} {\bibfield  {journal} {\bibinfo  {journal}
  {Physical Review E}\ }\textbf {\bibinfo {volume} {107}},\ \bibinfo {pages}
  {L062101} (\bibinfo {year} {2023})}\BibitemShut {NoStop}%
\bibitem [{\citenamefont {Zubkov}(1999)}]{Zubkov_1999}%
  \BibitemOpen
  \bibfield  {author} {\bibinfo {author} {\bibfnamefont {A.}~\bibnamefont
  {Zubkov}},\ }\href@noop {} {\bibfield  {journal} {\bibinfo  {journal} {Theory
  of Probability \& Its Applications}\ }\textbf {\bibinfo {volume} {43}},\
  \bibinfo {pages} {676} (\bibinfo {year} {1999})}\BibitemShut {NoStop}%
\bibitem [{\citenamefont {Nikitin}\ and\ \citenamefont
  {Belan}(2023)}]{nikitin2023choosing}%
  \BibitemOpen
  \bibfield  {author} {\bibinfo {author} {\bibfnamefont {I.}~\bibnamefont
  {Nikitin}}\ and\ \bibinfo {author} {\bibfnamefont {S.}~\bibnamefont
  {Belan}},\ }\href@noop {} {\bibfield  {journal} {\bibinfo  {journal} {arXiv
  preprint arXiv:2309.05877}\ } (\bibinfo {year} {2023})}\BibitemShut {NoStop}%
\bibitem [{\citenamefont {Pearson}(1916)}]{pearson1916ix}%
  \BibitemOpen
  \bibfield  {author} {\bibinfo {author} {\bibfnamefont {K.}~\bibnamefont
  {Pearson}},\ }\href@noop {} {\bibfield  {journal} {\bibinfo  {journal}
  {Philosophical Transactions of the Royal Society of London. Series A,
  Containing Papers of a Mathematical or Physical Character}\ }\textbf
  {\bibinfo {volume} {216}},\ \bibinfo {pages} {429} (\bibinfo {year}
  {1916})}\BibitemShut {NoStop}%
\bibitem [{\citenamefont {Sharma}\ and\ \citenamefont
  {Bhandari}(2015)}]{sharma2015skewness}%
  \BibitemOpen
  \bibfield  {author} {\bibinfo {author} {\bibfnamefont {R.}~\bibnamefont
  {Sharma}}\ and\ \bibinfo {author} {\bibfnamefont {R.}~\bibnamefont
  {Bhandari}},\ }\href@noop {} {\  (\bibinfo {year} {2015})}\BibitemShut
  {NoStop}%
\bibitem [{\citenamefont {Belan}(2018)}]{Belan_PRL_2018}%
  \BibitemOpen
  \bibfield  {author} {\bibinfo {author} {\bibfnamefont {S.}~\bibnamefont
  {Belan}},\ }\href@noop {} {\bibfield  {journal} {\bibinfo  {journal}
  {Physical Review Letters}\ }\textbf {\bibinfo {volume} {120}},\ \bibinfo
  {pages} {080601} (\bibinfo {year} {2018})}\BibitemShut {NoStop}%
\bibitem [{\citenamefont {Pal}\ and\ \citenamefont
  {Prasad}(2019{\natexlab{b}})}]{pal2019first}%
  \BibitemOpen
  \bibfield  {author} {\bibinfo {author} {\bibfnamefont {A.}~\bibnamefont
  {Pal}}\ and\ \bibinfo {author} {\bibfnamefont {V.}~\bibnamefont {Prasad}},\
  }\href@noop {} {\bibfield  {journal} {\bibinfo  {journal} {Physical Review
  E}\ }\textbf {\bibinfo {volume} {99}},\ \bibinfo {pages} {032123} (\bibinfo
  {year} {2019}{\natexlab{b}})}\BibitemShut {NoStop}%
\bibitem [{\citenamefont {Singh}\ \emph
  {et~al.}(2023{\natexlab{a}})\citenamefont {Singh}, \citenamefont {Sandev},\
  and\ \citenamefont {Singh}}]{singh2023escape}%
  \BibitemOpen
  \bibfield  {author} {\bibinfo {author} {\bibfnamefont {R.}~\bibnamefont
  {Singh}}, \bibinfo {author} {\bibfnamefont {T.}~\bibnamefont {Sandev}}, \
  and\ \bibinfo {author} {\bibfnamefont {S.}~\bibnamefont {Singh}},\
  }\href@noop {} {\bibfield  {journal} {\bibinfo  {journal} {arXiv preprint
  arXiv:2305.01601}\ } (\bibinfo {year} {2023}{\natexlab{a}})}\BibitemShut
  {NoStop}%
\bibitem [{\citenamefont {Chechkin}\ and\ \citenamefont
  {Sokolov}(2018)}]{chechkin2018random}%
  \BibitemOpen
  \bibfield  {author} {\bibinfo {author} {\bibfnamefont {A.}~\bibnamefont
  {Chechkin}}\ and\ \bibinfo {author} {\bibfnamefont {I.}~\bibnamefont
  {Sokolov}},\ }\href@noop {} {\bibfield  {journal} {\bibinfo  {journal}
  {Physical review letters}\ }\textbf {\bibinfo {volume} {121}},\ \bibinfo
  {pages} {050601} (\bibinfo {year} {2018})}\BibitemShut {NoStop}%
\bibitem [{\citenamefont {Singh}\ \emph
  {et~al.}(2023{\natexlab{b}})\citenamefont {Singh}, \citenamefont {Sandev},\
  and\ \citenamefont {Singh}}]{singh2023bernoulli}%
  \BibitemOpen
  \bibfield  {author} {\bibinfo {author} {\bibfnamefont {R.}~\bibnamefont
  {Singh}}, \bibinfo {author} {\bibfnamefont {T.}~\bibnamefont {Sandev}}, \
  and\ \bibinfo {author} {\bibfnamefont {S.}~\bibnamefont {Singh}},\
  }\href@noop {} {\bibfield  {journal} {\bibinfo  {journal} {Physical Review
  E}\ }\textbf {\bibinfo {volume} {108}},\ \bibinfo {pages} {L052106} (\bibinfo
  {year} {2023}{\natexlab{b}})}\BibitemShut {NoStop}%
\end{thebibliography}%

\end{document}